\documentclass[12pt]{iopart}

\begin{document}
\title{Diffeomorphism covariant star products and noncommutative
gravity}
\author{D V Vassilevich}
\address{CMCC, Universidade Federal do ABC, Santo Andr\'e, SP, Brazil}
\address{Department of Theoretical Physics, St Petersburg State University,
St Petersburg, Russia}
\ead{dvassil@gmail.com}
\begin{abstract}The use of a diffeomorphism covariant star product enables
us to construct diffeomorphism invariant gravities on noncommutative 
symplectic manifolds without twisting the symmetries. 
As an example, we construct noncommutative deformations
of all two-dimensional dilaton gravity models thus overcoming some difficulties
of earlier approaches. One of such models appears to be
integrable. We find all classical solutions of this model and discuss
their properties.
\end{abstract}
\section{Introduction}\label{sec-in}
The diffeomorphism invariance is one of the most important features of
any gravity theory. Despite recent advances in noncommutative gravity 
\cite{Szabo:2006wx} there is still no unique and totally satisfactory
way to realize the full diffeomorphism group on a noncommutative
manifold. One can use the Seiberg-Witten approach \cite{SW} which
reduces all symmetries, including the diffeomorphisms, to the commutative
ones at the expense of a non-linear field redefinition. However,
calculations beyond the leading order in the noncommutativity
parameter are hardly possible in this approach, 
see, e.g, \cite{SWmore}. Another way
to extend the diffeomorphism transformations to noncommutative
spaces is to make their action twisted \cite{twidif}. One can
construct a full twisted-invariant\footnote{The first symmetry to be
twisted was the global Poincare symmetry \cite{Masud}.} 
noncommutative gravity action having
just the right number of symmetries. However, the twisted symmetries are
not {\it bona fide} physical symmetries. One cannot use them to gauge
away any degrees of freedom.

Here we develop a different approach to the diffeomorphism invariance
on noncommutative spaces. The star product we use is a particular case of
the geometric construction by Fedosov \cite{Fedosov} suggested in
\cite{Batalin:1990yk}. This star product is diffeomorphism covariant. 
As we show, the tensor algebra built up with this star product
has many nice properties and is suitable for the construction of gravity
theories on noncommutative manifolds. As an example, we consider dilaton
gravities in two dimensions and show that all of them have fully
diffeomorphism invariant noncommutative counterparts. One of these models
(a conformally transformed Witten black hole model) appears to be
classically integrable in the noncommutative case. We construct
all solutions of this model and discuss briefly their properties.  
\section{The star product}\label{sec-star}
Let us suppose that the space-time $\mathcal{M}$ is a symplectic
manifold. That is, $\mathcal{M}$ is equipped with a closed 
non-degenerate two-form
$\omega$. In a local coordinate system this implies that
\begin{equation}
\partial_\mu\omega_{\nu\rho}+\partial_\rho\omega_{\mu\nu}
+\partial_\nu\omega_{\rho\mu}=0\,.
\label{c2f}
\end{equation}
The inverse of $\omega_{\mu\nu}$, $\omega^{\nu\rho}$,
is  defined through the equation
\begin{equation}
\omega_{\mu\nu}\omega^{\nu\rho}=\delta_\mu^\rho\,.
\label{invo}
\end{equation}
$\omega^{\nu\rho}$ is a Poisson bivector.
It satisfies the Jacobi identities 
as a consequence of (\ref{c2f}).

Let us choose a Christoffel symbol
on $\mathcal{M}$ such that the symplectic form is covariantly constant,
\begin{equation}
\nabla_{\mu}\omega_{\nu\rho}=
\partial_\mu \omega_{\nu\rho}-\Gamma_{\mu\nu}^{\sigma}\omega_{\sigma\rho}
-\Gamma_{\mu\rho}^\sigma \omega_{\nu\sigma}=0.\label{cco}
\end{equation}
Thus $\mathcal{M}$ is a Fedosov manifold \cite{GRS}. Let us suppose
that the connection $\Gamma$ is flat, i.e. the curvature 
tensor\footnote{In \cite{Batalin:1990yk} the tensor (\ref{cur})
is called the Riemannian curvature. We shall avoid this terminology
since no Riemannian structure (metric) is assumed on $\mathcal{M}$.} 
\begin{equation}
R^{\mu}_{\ \ \nu\rho\sigma}=
\partial_{\rho}\Gamma_{\sigma\nu}^{\mu}
-\partial_{\sigma}\Gamma_{\rho\nu}^{\mu}
+\Gamma_{\sigma\nu}^{\lambda}\Gamma^\mu_{\rho\lambda}
-\Gamma_{\rho\nu}^{\lambda}\Gamma^\mu_{\sigma\lambda}
\label{cur}
\end{equation}
vanishes. We also suppose that the connection is symmetric,
$\Gamma_{\mu\nu}^\rho=\Gamma_{\nu\mu}^\rho$, which implies that
the torsion vanishes. Due to these two assumptions, the covariant
derivatives commute.

We can now define a star product
\begin{eqnarray}
f\star g &=& f \exp\left( \overleftarrow{\nabla}_\mu
\frac i2 \omega^{\mu\nu} \overrightarrow{\nabla}_\nu
\right)g \nonumber\\
&=&\sum_{n=0}^{\infty} \frac 1{n!}\left( \frac i2\right)^n
\omega^{\mu_1\nu_1}\dots \omega^{\mu_n\nu_n}
(\nabla_{\mu_1}\dots \nabla_{\mu_n} f)
(\nabla_{\nu_1}\dots \nabla_{\nu_n} g)\,.
\label{star}
\end{eqnarray}
The algebra of smooth functions on $\mathcal{M}$ with this multiplication will
be denoted $\mathcal{A}_\omega$. This is an associative 
noncommutative algebra. The product (\ref{star}) solves the deformation 
quantization problem \cite{BFFLS} in the sense that for any two 
scalar functions $f$ and $g$ on $\mathcal{M}$
\begin{equation}
f\star g =f\cdot g + \frac i2 \omega^{\mu\nu}\partial_\mu f\cdot
\partial_\nu g +O\left((\omega^{\mu\nu})^2\right),\label{exstar}
\end{equation}
i.e., the linear in $\omega^{\mu\nu}$ term reproduces the Poisson
bracket between $f$ and $g$. For another diffeomorphism covariant
star product introduced in \cite{cor} only e weaker property holds.
Namely, the Poisson bracket is given by the linear order of the
star-commutator.

Note, that the use of non-flat connections leads to non-associative
star products \cite{Harikumar:2006xf,Chaichian:2006wt}.
In generic star products \cite{Kontsevich} partial derivatives of 
$\omega^{\mu\nu}$ appear. Such objects are, however, non-covariant.
It is therefore doubtful that they may be used to build up covariant
star products.

The star product (\ref{star}) is a particular case of the Fedosov construction
\cite{Fedosov}, which was proposed in \cite{Batalin:1990yk}. Let us formulate 
some basic properties of the product (\ref{star}). Obviously,
this product may be extended from functions to arbitrary tensors. Therefore,
in the formulae below $f$ and $g$ are tensor fields. First we 
observe that due to (\ref{cco})
\begin{equation}
\omega^{\mu\nu}\star f=\omega^{\mu\nu}\cdot f \label{ost}\,,
\end{equation}
i.e., $\omega^{\mu\nu}$ belongs to the center of corresponding
commutator algebra. 
One can also see that $\nabla_\mu$ is a derivation on $\mathcal{A}_\omega$,
\begin{equation}
\nabla_\mu (f\star g)=(\nabla_\mu f)\star g + f \star (\nabla_\mu g).
\label{der}
\end{equation} 
The product (\ref{star}) is hermitian,
\begin{equation}
\overline{(f\star g)}=\bar g \star \bar f\,,\label{her}
\end{equation}
where the bar denotes complex conjugation. 

In the context of this work, the most important property of the star product
(\ref{star}) is the {\it diffeomorphism covariance}. Let $f\to f'$ be a 
diffeomorphism transformation, then
\begin{equation}
(f\star g)'=f'\star' g' \,,\label{difco}
\end{equation}
where $\star'$ is given by the formula (\ref{star}) where $\omega^{\mu\nu}$
and the connection are transformed under the diffeomorphism in the standard
way (as in the commutative geometry). 
The star product preserves its' form under the action of commutative
diffeomorphisms and is, therefore, a diffeomorphism scalar.
Again, $f$ and $g$ in (\ref{difco}) 
may be tensors of any rank. 

The star multiplication commutes with lowering/raising indices with 
$\omega_{\mu\nu}$ and $\omega^{\mu\nu}$, respectively. For example,
$(f_\nu\omega^{\nu\mu}) \star g = f^\mu \star g
=(f_\nu \star g)\omega^{\nu\mu}$.

There is a natural integration measure 
\cite{Fedosov,Batalin:1990yk,Felder:2000nc}
\begin{equation}
d\mu (x)=(\det (\omega^{\mu\nu}))^{-\frac 12}\, dx\,.
\label{intme}
\end{equation}
It is easy to check that with respect to this measure the star product
of tensors is closed provided all indices are contracted in pairs 
(i.e., if the integrand is a diffeomorphism scalar),
\begin{equation}
\int_{\mathcal{M}} d\mu (x) f_{\mu\nu\dots\rho}\star g^{\mu\nu\dots\rho}=
\int_{\mathcal{M}} d\mu (x) f_{\mu\nu\dots\rho}\cdot g^{\mu\nu\dots\rho}\,.
\label{clo}
\end{equation}
This equation also implies that contraction of all indices and integration
with the measure (\ref{intme}) is a trace on the star-tensor algebra
over $\mathcal{M}$. Given nice properties of the tensor algebra one can
expect that this approach will also shed a new light onto the problem
of construction of Poisson structures and star products on
the exterior algebras (differential forms), see \cite{forms}. 

In the Riemannian geometry a torsionless connection is uniquely fixed
by the condition of covariant constancy of the metric. In the symplectic
geometry, the condition (\ref{cco}) does not fix the connection uniquely, 
even if one requires that the torsion and the curvature vanish.
As noted e.g. in \cite{Batalin:1990yk}, a flat torsionless symplectic
connection trivializes, $\nabla_\mu = \partial_\mu$, in a Darboux
coordinate system. The Darboux coordinates are defined up to a 
symplectomorphysm, which is a diffeomorphism preserving $\omega_{\mu\nu}$.
Therefore, to fix a star product one has to fix a symplectic form
$\omega_{\mu\nu}$ {\it and} a symplectomorphism.
\section{Noncommutative gravity in two dimensions}
Let us first consider generic two-dimensional dilaton gravity
on a commutative space \cite{Grumiller:2002nm}. The Euclidean
first-order action reads \cite{Bergamin:2004pn}
\begin{eqnarray}
&&S_{\rm c}=\int d^2x \,\epsilon^{\mu\nu} \left[
\bar Y (\partial_\mu e_\nu -i\rho_\mu e_\nu)+
Y (\partial_\mu \bar e_\nu + i\rho_\mu \bar e_\nu)\right.
\nonumber\\
&&\qquad\qquad\qquad\qquad\qquad \left.
+\Phi \partial_\mu\rho_\nu +
iV(\Phi)\bar e_\mu e_\nu \right]\,. \label{scom}
\end{eqnarray}
We use complex fields
\begin{eqnarray}
&Y=\frac 1{\sqrt{2}} (Y^1+iY^2),\qquad & 
\bar Y=\frac 1{\sqrt{2}} (Y^1-iY^2),\nonumber\\
&e=\frac 1{\sqrt{2}} (e^1+ie^2),\qquad & 
\bar e=\frac 1{\sqrt{2}} (e^1-ie^2),\label{cfield}
\end{eqnarray}
where the superscript is a $U(1)$ (Euclidean Lorentz) index. 
$\epsilon^{\mu\nu}$ is the Levi-Civita symbol, $\epsilon^{12}=
-\epsilon^{21}=1$.
$e_\mu$ is the zweibein, and $\rho_\mu$ is the spin connection.
$\Phi$ is the dilaton, and $Y$ and $\bar Y$ are auxiliary fields
which generate the torsion constraint. $V(\Phi)$ is an arbitrary
function of the dilaton, which defines a particular model within the
family. Most general dilaton gravity actions contain also a 
term $U(\Phi)\bar YY$. Such a term can be removed by a conformal
redefinition of the metric\footnote{This conformal redefinition
does not provide a full equivalence between models even classically
since the corresponding conformal transformation is in general valid
only locally. Here we ignore this subtlety and consider the theories
with vanishing $U(\Phi)$.}. An extensive list of physically relevant
potentials $V$ and $U$ can be found in 
\cite{Grumiller:2002nm,Grumiller:2006rc}.

A noncommutative extension of the action (\ref{scom}) reads
\begin{eqnarray}
&&S_{\rm nc}=\int d\mu(x) \,\omega^{\mu\nu} \star \left[
\bar Y\star (\nabla_\mu e_\nu -i\rho_\mu\star e_\nu)+
 (\nabla_\mu \bar e_\nu - i \bar e_\mu \star \rho_\nu )\star Y \right.
\nonumber\\
&&\qquad\qquad\qquad\qquad \left.
+\Phi\star (\nabla_\mu\rho_\nu - i\rho_\mu \star \rho_\nu) +
i\bar e_\mu \star V_\star (\Phi) \star e_\nu \right]\,. \label{snc}
\end{eqnarray}
This action is invariant under the following noncommutative
$U(1)_\star$ gauge transformations
\begin{eqnarray}
&\delta Y=i\lambda \star Y,\qquad & 
\delta \bar Y=-i\bar Y \star \lambda,\nonumber\\
&\delta e_\mu =i\lambda \star e_\mu ,\qquad & 
\delta \bar e_\mu =-i\bar e_\mu \star \lambda,\nonumber\\
&\delta \rho_\mu =\partial_\mu\lambda -i[\rho_\mu,\lambda]_\star ,
\qquad & \delta \Phi=i[\lambda,\Phi]_\star\,, \label{gtra}
\end{eqnarray}
which will be treated as a noncommutative extension of the Euclidean
Lorentz symmetry. In the equations above, $[\cdot ,\cdot ]_\star$
denotes the star-commutator, $[\lambda,\Phi]_\star \equiv
\lambda \star \Phi - \Phi \star \lambda$. 
The Poisson bivector $\omega^{\mu\nu}$ is invariant under the $U(1)_\star$
transformations. This is consistent with the general rule, that $U(1)$
invariants transform in the noncommutative case through star-commutators,
and $[\omega^{\mu\nu},\lambda]_\star =0$ due to (\ref{ost}).
The invariance of the action (\ref{snc}) with respect to diffeomorphisms
is ensured by the diffeomorphism covariance of the star product.

In two dimensions any antisymmetric tensor is proportional to the Levi-Civita
symbol, 
\begin{equation}
\omega^{\mu\nu}=B(x)\epsilon^{\mu\nu}. \label{2Dom}
\end{equation}
Therefore, 
${\mathrm {det}}\, (\omega^{\mu\nu})=B^2$,
$d\mu(x)\omega^{\mu\nu}=d^2x \epsilon^{\mu\nu}$,  
and the action (\ref{snc})
indeed reproduces (\ref{scom}) in the commutative limit\footnote{Strictly
speaking, the limit $\omega^{\mu\nu}\to 0$ does not exist since 
$\omega_{\mu\nu}$ diverges. Instead of taking  $\omega^{\mu\nu}\to 0$,
one has to replace $\omega^{\mu\nu}$ by $\alpha\omega^{\mu\nu}$ in the
star product (\ref{star}) and then take $\alpha\to 0$.}. Although the
integration measure diverges in the commutative limit, this divergence
is cancelled by the behavior of $\omega^{\mu\nu}$ in the Lagrangian.
Non-singularity of the action in the commutative limit can serve as
a criterion to select suitable noncommutative actions in four dimensions.

Note, that the deformation (\ref{snc}) of (\ref{scom}) is ``fairly unique''.
This means the following. Of course, one has to make a choice, whether
the noncommutative gauge transformations act on $Y$ from the left, or
from the right. If they act from the left, as in (\ref{gtra}), the
gauge transformation for $\bar e_\mu$ is defined uniquely since we are
going to couple $\bar e_\mu$ to $Y$. The transformations for $\bar Y$
and $e_\mu$ then follow by complex conjugation. The transformation
rules of $\rho_\mu$ and $\Phi$ and the action (\ref{snc}) are then fixed
uniquely by requiring that the commutative limit is (\ref{scom}), and that
the action is real and gauge invariant provided we do not include any terms 
containing products of $Y$ and $\bar Y$. This is in contrast to the term
$U(\Phi)\bar Y Y$ which we discussed briefly below eq. (\ref{cfield}).
Any interaction of the form 
$i\sum_a \bar e_\mu \star W_\star^{[a]}(\Phi)\star e_\nu
\star \bar Y \star \tilde W_\star^{[a]} (\Phi) \star Y$ 
with the only restriction
$\sum_a W^{[a]}(\Phi)\tilde W^{[a]}(\Phi)=U(\Phi)$ will (after the
integration over $\mathcal{M}$) be real, 
gauge-invariant, and possess a correct commutative limit. A physical 
interpretation of this enormous ambiguity remains unclear. To avoid this
ambiguity we shall not consider any interactions containing both $\bar Y$
and $Y$.

We like to stress, that in this approach one can construct a deformation
of {\it any} $2D$ dilaton gravity model in such a way that the deformed
model is invariant under diffeomorphisms and deformed 
Lorentz transformations.
Previously, a noncommutative deformation with untwisted symmetries
was constructed only for the Jackiw-Teitelboim \cite{JT} 
model (linear $V(\Phi)$,
$U(\Phi)=0$) by using its equivalence to a BF model with Yang-Mills
type symmetries \cite{Cacciatori:2002ib}. (This model appeared to be
even quantum integrable \cite{Vassilevich:2004ym}).
Later it was demonstrated, that
one cannot add higher order terms to the linear
potential of the model \cite{Cacciatori:2002ib} and preserve the
number of symmetries in a noncommutative gravity theory in two dimensions
\cite{Vassilevich:2006uv}.

There is an interesting relation to the twisted-symmetric models. By taking a
constant $\omega^{\mu\nu}$ and ``gauge fixing'' the connection in $\nabla_\mu$
to zero one arrives at twisted diffeomorphism invariant gravity in $2D$
(cf. \cite{Balachandran:2006qg}). This is in parallel to the observation
made in \cite{Vassilevich:2007jg} in the context of the Yang-Mills
symmetries. By fixing a gauge in the gauge covariant star product one 
obtains a twisted-symmetric Yang-Mills theory \cite{twist}. 

Let us consider a noncommutative version of the Witten black hole 
\cite{Wbh}. After a conformal redefinition of the metric \cite{CGHS}
in the commutative case one obtains the action (\ref{scom}) with
a constant potential
\begin{equation}
V(\Phi)=\Lambda\,. \label{cV}
\end{equation}
This model is almost trivial since it describes the flat metric only.
In the noncommutative case,
the equations of motion following from the action (\ref{snc}) with 
the potential (\ref{cV}) read
\begin{eqnarray}
&& \epsilon^{\mu\nu}(\nabla_\mu \rho_\nu -i \rho_\mu \star \rho_\nu)=0,
\label{Peq}\\
&& \epsilon^{\mu\nu}(\nabla_\mu e_\nu - i \rho_\mu \star e_\nu )=0,
\label{bYeq}\\
&& \epsilon^{\mu\nu}(\nabla_\mu \bar e_\nu - i\bar e_\mu \star \rho_\nu)=0,
\label{Yeq}\\
&&\nabla_\nu \Phi +i[\Phi,\rho_\nu]_\star  
-ie_\nu \star \bar Y + iY\star \bar e_\nu =0,\label{req}\\
&&\nabla_\nu \bar Y + i \bar Y \star \rho_\nu - i \Lambda \bar e_\nu =0,
\label{eeq}\\
&&\nabla_\nu Y -i \rho_\nu \star Y +i \Lambda e_\nu=0,\label{beeq}
\end{eqnarray}

Note, that in the equations (\ref{Peq}) - (\ref{beeq}) one can replace
the covariant derivatives
$\nabla_\mu$ by the partial derivatives $\partial_\mu$ (except for the
covariant derivatives hidden in the star product). The reason
is that these derivatives either act on scalars or appear contracted with
the Levi-Civita symbol, as, for example, 
$\epsilon^{\mu\nu}\nabla_\mu\rho_\nu$ in (\ref{Peq}).

In all commutative $2D$ dilaton gravity theories there is a quantity
$\mathcal{C}(\Phi, \bar Y, Y)$ which is absolutely conserved,
$\partial_\mu \mathcal{C}(\Phi, \bar Y, Y)=0$, due to the equations of motion.
The existence of this quantity is essential for the classical integrability
of dilaton gravities. For example, for a constant dilaton potential $V$ given
by (\ref{cV}) the conserved quantity reads $\mathcal{C}=Y\bar Y +\Lambda \Phi$.

Let us try to define a similar quantity in the
noncommutative case. This can be done in the same way as in commutative
Euclidean theories \cite{Bergamin:2004pn}. One only has to fix properly 
the order of
multipliers. Let us multiply eq.\ (\ref{eeq}) by $Y$ from the left and
add to the equation (\ref{beeq}) multiplied by $\bar Y$ from the right.
Then, use eq.\ (\ref{req}) to get rid of the terms containing $e$ and $\bar e$.
We have
\begin{equation}
(\nabla_\mu - i \, {\mathrm{ad}}_\star\, \rho_\mu) 
(Y \star\bar Y +\Lambda \Phi)=0
\,,\label{cons}
\end{equation}
where $({\mathrm{ad}}_\star\, a) b=[a,b]_\star$ is the adjoint action.
In contrast to the commutative case, equation (\ref{cons}) contains a
$U(1)_\star$ covariant derivative. The model is nevertheless classically
integrable. The equation (\ref{Peq}) yields that $\rho_\mu$ is
a trivial $U(1)_\star$ connection at least locally, i.e.,
\begin{equation}
\rho_\mu = iu\star \nabla_\mu u^{-1}\,,\label{ru}
\end{equation}
where $\bar u=u^{-1}$ and $u\star u^{-1}=1$. Let us introduce gauge transformed
fields
\begin{eqnarray}
&Y=u\star Y^u,\qquad & e_\mu =u\star e_\mu^u,\nonumber\\
&\bar Y =\bar Y^u\star u^{-1},\qquad &\bar e_\mu =\bar e_\mu^u \star u^{-1},
\nonumber\\
&\Phi=u\star \Phi^{u}\star u^{-1}.  &\label{ufield}
\end{eqnarray}
Next, let us substitute the fields (\ref{ufield}) into eqs.\ (\ref{bYeq}) -
(\ref{beeq}). The equations still have the same form in terms 
of transformed the fields
$\{ Y^u,\bar Y^u, e_\mu^u,\bar e_\mu^u,\Phi^u \}$, 
except that $\rho_\mu$ disappears.
One then easily finds a general solution
\begin{equation}
e_\mu^u=\nabla_\mu E,\qquad Y^u =-i\Lambda E,\qquad 
\Phi^u=b-\frac 1\Lambda Y^u\star \bar Y^u\,,\label{sol}
\end{equation}
where $E$ is an arbitrary complex function, $b$ is an arbitrary
real constant. The solutions for $\bar e_\mu^u$ and
$\bar Y^u$ are given by complex conjugation.

The solution depends on three arbitrary real functions (one parametrizes
$u$, and the other two are the real and imaginary parts of $E$, respectively).
This corresponds to the presence of three local symmetries of the action
(\ref{snc}) (two diffeomorphisms and one $U(1)_\star$).

One can define an $U(1)_\star$ (Lorentz) invariant tensor
\begin{equation}
g_{\mu\nu}=\frac 12 (\bar e_\mu \star e_\nu +
\bar e_\nu \star e_\mu),\label{gmn}
\end{equation}
which may be identified with the Riemannian metric on $\mathcal{M}$.
The line element $(ds)^2=g_{\mu\nu}dx^\mu dx^\nu$ is diffeomorphism invariant
in the standard sense, i.e., it does not change under the coordinate 
transformations.

The solution (\ref{ru}), (\ref{sol}) is flat. The connection $\rho_\mu$ is
a gauge-trivial one, and the zweibein $e_\mu^u$ can be reduced, at least
locally, to the unit one. This can be done, e.g., by choosing the coordinates
$x^1={\mathrm Re}\, E$, $x^2={\mathrm Im}\, E$. However, neither 
$\omega^{\mu\nu}$ needs be a constant, nor $\nabla_\mu$ needs be trivial
in precisely the same coordinate system where the zweibein is trivial. 
Consequently, the metric
(\ref{gmn}) {\it need not be trivial} since it is constructed by using the
star product. In general, $g_{\mu\nu}$ cannot be reduced to the unit one
by choosing a suitable coordinate system even locally.
\section{Conclusions}\label{sec-con}
In this paper we considered a diffeomorphism covariant star product
on a symplectic manifold and studied the properties of corresponding 
tensor algebra. We constructed noncommutative diffeomorphism invariant
deformations of all dilaton gravities in two dimensions thus overcoming
some difficulties of earlier approaches. For the simplest model with
a constant dilaton potential we were able to find all classical solutions.
Although the solutions correspond to a flat zweibein $e_\mu$
and to a flat spin-connection $\rho_\mu$, the metric need not
be flat.

There are many possible extensions of the results reported above.
The most immediate one is to consider the dilaton potentials other
than the constant one (\ref{cV}). It is not obvious whether corresponding
noncommutative gravities will be integrable. An extension to four-dimensional
gravities also looks rather straightforward. Although $SO(1,3)$ and
$SO(4)$ do not close on noncommutative spaces, one can either work in the 
metric formalism thus avoiding Lorentz transformations, or add a trivial
Lorentz or $SO(4)$ connection to the covariant derivative $\nabla$
in the star product.

The restriction to symplectic
manifolds may be weakened. One can consider instead of symplectic manifolds
regular Poisson 
manifolds where the Fedosov construction \cite{Fedosov} also works
well.

In the model we considered in this paper the symplectic geometry plays
the role of ``external conditions'' which were not restricted by
any equations of motion.
It would be interesting to make dynamical fields out of the symplectic
structure $\omega_{\mu\nu}$ and the symplectomorphism which define
the star product (see the discussion at the end of sec.\ \ref{sec-star}).
In this respect we like to mention the approach of Pinzul and Stern 
\cite{Pinzul:2007bk}.
\ack
This work is supported in part by the CNPq and by the project
RNP 2.1.1/1575. I am grateful to the Erwin Schr\"{o}dinger International
Institute for Mathematical Physics for
warm hospitality in Vienna where this work was completed.

\end{document}